\documentclass[conference]{IEEEtran}
\IEEEoverridecommandlockouts
\usepackage{fancyhdr}
\usepackage{cite}
\usepackage{amsmath,amssymb,amsfonts}
\usepackage{algorithmic}
\usepackage{graphicx}
\usepackage{textcomp}
\usepackage{xcolor}
\def\BibTeX{{\rm B\kern-.05em{\sc i\kern-.025em b}\kern-.08em
    T\kern-.1667em\lower.7ex\hbox{E}\kern-.125emX}}

\usepackage[belowskip=-5pt,aboveskip=-2pt]{caption}

\usepackage{algorithm2e}

\SetKwInput{KwInput}{Input}                
\SetKwInput{KwOutput}{Output}              
\SetKwInput{KwTrigger}{@Trigger}              
\usepackage[numbers,sort&compress]{natbib}

\begin{document}

\title{Collaborative adversary nodes learning on the logs of IoT devices in an IoT network}

\author{
	
	\IEEEauthorblockN{
		Sandhya Aneja\IEEEauthorrefmark{1},
		Melanie Ang Xuan En\IEEEauthorrefmark{1}, 
		Nagender Aneja\IEEEauthorrefmark{2} 
	}
	\IEEEauthorblockA{ \IEEEauthorrefmark{1}Faculty of Integrated Technologies, Universiti Brunei Darussalam \\
		\IEEEauthorrefmark{2}School of Digital Science, Universiti Brunei Darussalam,
	}
	Email: \{sandhya.aneja, 16b4003, nagender.aneja\}@ubd.edu.bn
}

\maketitle
\thispagestyle{fancy}

\begin{abstract}
Artificial Intelligence (AI) development has encouraged many new research areas, including AI-enabled Internet of Things (IoT) network. AI analytics and intelligent paradigms greatly improve learning efficiency and accuracy. Applying these learning paradigms to network scenarios provide technical advantages of new networking solutions. In this paper, we propose an improved approach for IoT security from data perspective.  The network traffic of IoT devices can be analyzed using AI techniques. The Adversary Learning (AdLIoTLog) model is proposed using Recurrent Neural Network (RNN) with attention mechanism on sequences of network events in the network traffic. We define network events as a  sequence of the time series packets of protocols captured in the log. We have considered different packets TCP packets, UDP packets, and HTTP packets in the network log to make the algorithm robust. The distributed IoT devices can collaborate to  cripple our world which is extending to Internet of Intelligence. The time series packets are converted into structured data by removing noise and adding timestamps. The resulting data set is trained by RNN and can detect the node pairs collaborating with each other. We used the BLEU score to evaluate the model performance. Our results show that the predicting performance of the AdLIoTLog model trained by our method degrades by 3-4\% in the presence of attack in comparison to the scenario when the network is not under attack. AdLIoTLog can detect adversaries because when adversaries are present the model gets duped by the collaborative events and therefore predicts the next event with a biased event rather than a benign event.  We conclude that AI can provision ubiquitous learning for the new generation of Internet of Things.
\end{abstract}


\vspace{-.3cm}

\section{Introduction}
\begin{figure}[!bhtp]
	\centering
	\includegraphics[width=2.5in, height=1.5in]{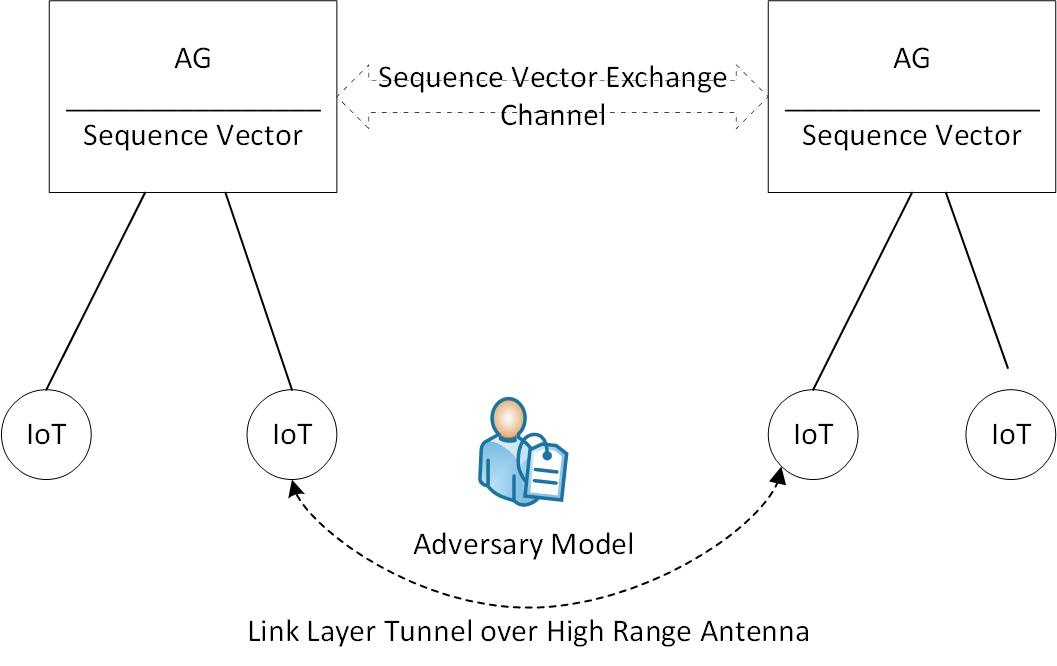}
	\caption{Deep Adversary Architecture}
	\label{fig:DAA}
\end{figure}

The Internet of Things ((IoT) devices are resource-constrained low power devices collecting a large volume of data for IoT applications in healthcare, retail,  transportation and manufacturing.  The data collected through IoT applications is valuable and huge. Many malwares, and botnets have been observed to compromise the IoT devices  by leveraging the vulnerabilities like default passwords.  The attackers can   affect the physical state of the device once the device  is compromised \cite{hou2019survey, hassija2020security}.
IoT devices are vulnerable to the scenario where devices connected to a gateway can collaborate to mislead the smart decision of the IoT network. One of the scenarios is wherein IoT devices in two different LANs or locations can collaborate using a high transmission antenna to exchange data say temperature, pressure, and humidity. The collaborating IoT devices can then upload the distant location data to the server with its own location which can cripple the system due to high temperature maliciously reported as low temperature. Adversary Learning (AdLIoTLog) framework collects log files from the various application gateways  \cite{robyns2017noncooperative} and apply deep learning \cite{shen2018tiresias, sweet2019veracity} to detect collaborating nodes connected through high transmission power channel for adversary behavior to other nodes shown in Figure \ref{fig:DAA}. 

While capturing data through IoT, metadata can also be captured to apply AI techniques for IoT network security.  Traditional AI techniques were about centralized data. In another AI paradigm called as federated learning (FL) model is trained from distributed systems over the cloud. Here interesting observation for FL is that the learned model over distributed systems can be secured like other encrypted numbers communicated over the Internet  {\cite{nguyen2021federated}}. A simple/low-complexity resource allocation algorithm is proposed for a wireless network to support multiple FL groups~\cite{vu2021does}. IoT devices may be compromised. We propose in this paper to analyze network traffic logs of IoT devices distributed in a network behind the application gateways. This network traffic logged at application gateways can be used to identify compromised devices as well as collaborative adversaries.

The IoT devices log comprises the chronological events of the packets of these protocols. The packets include port numbers, IP addresses, sequence numbers, flags, checksum, window size and domain names. For example, domains such as example.com, example.net, and example.org are frequently requested by Amazon Echo; sub-domains of hp.com and hpeprint.com are seen in DNS queries from the HP printer~\cite{sivanathan2018classifying}. Due to this complex structure of IoT log data, it is complicated to analyze logs for any information. 

A similar method \cite{miettinen2017iot, aneja2018iot} is presented at the application gateway to authenticate the IoT device by analyzing the 212 features like TCP src port and TCP dst port from the packet headers of IoT devices  in the logged network traffic.

Log analysis using the Recurrent Neural Network (RNN) method has been \cite{wu2019bigdata, shen2018tiresias} studied to predict future events. 
In ns-2 network simulator, two network scenarios were set up to generate data for the proposed study. The trace files log the sequences of network events of the nodes comprising of different types of protocols packets. The first scenario was a network without any adversary node while the second scenario was the network with collaborating adversary nodes which were connected through a link layer tunnel as hidden channel for adversary behavior to other nodes.

The Adversary Learning model degrades by 3-4\% in the presence of attack in comparison to the scenario when the network is not under attack. Model was found more robust for UDP packets in comparison to TCP and HTTP packets. A network protocol fixes the packet format in the network traffic of the devices. We observed that the Recurrent Neural Network models - LSTM, GRU etc. are learned with less execution time and better predicting for network problems in addition to language translation, emotion detection, and fake news detection problems.   Our contributions are as follows:

\begin{enumerate}
	\item  Collaborative adversary events detection was found effective using RNN.
	
	\item The network simulator ns-2 trace files generated for collaborative attack dataset and further interfaced in PyTorch for AI analytics using RNN model.
	
\end{enumerate}

\section{Gated Recurrent Neural Network- Sequence-to-Sequence Model} \label{rnn}.

We take the RNN-based GRU model \cite{bahdanau2014neural}  for the problem of  anomaly detection in an IoT Network.  Assume that the IoT network log vocabulary can be expressed as input network event  and predicted network event  using sequences   $x={x_1,x_2, \ldots, x_{|x|} }$ and $y={y_1,y_2, \ldots, y_{|y|} }$ respectively. LSTM and GRU create internal gates to regulate the information. These gates can learn the important data and can pass the relevant information until long chain. The core of the GRU is composed 
of encoder-decoder network which 
consists of three parts: (i) Encoder (ii) Attention Context Vector (iii) Decoder.

{\bf Encoder:} An encoder is a stack of many recurrent units where each accepts an element of the input network event sequence say $x={x_1,x_2, \ldots, x_{|x|} }$. 
The hidden states $h_t = f(W^{(hh)} h_{t-1}~+~W^{(hx)} x_t)$ are computed with the help of current input, previous state, and weights of the network. This is the final hidden state of the encoder. 

{\bf Attention:} The context vector aims to encapsulate input network event sequence information to assist the prediction of output network event sequence by a decoder. Context vector acts as an initial hidden state for the decoder. The context vector $c_p$ is computed as in Equation 	$c_{p} = \Sigma^{|x|}_{t=1} {\alpha_{tp}}h_{t}$  where $\alpha_{tp} = \frac {exp(r_{tp})}{\Sigma^{|x|}_{t=1}exp(r_{tp})} $ and $r_{tp} = v^{T}_{a}tanh(W^{(ss)}{s_{p-1}}+ W^{(hh)}h_{t-1})$ with the help of previous hidden state $h_{t-1}$, previous state $s_{p-1}$, and weights of the network normalized over the source sequence.

{\bf Decoder: } A decoder also comprises of recurrent units cells wherein each cell predicts an element of the output network event at a time step. Each recurrent unit cell accepts the previous target state $y_{p-1}$ and source context vector $c_p$ to produce output element $s_p = f(W^{(ss)}. s_{p-1} + ~W^{(sy)} y_{p-1} + ~W^{(sc)} c_p) $ and next  target hidden state.

The target hidden state $t_j = f(W^{(ss)}. s_{j} + ~W^{(sy)} y_{j-1} + ~W^{(sc)} c_j)$ is computed using the previous hidden state. The Probability distribution $P_j = softmax(W^s . t_j)$ over all elements of network event in target vocabulary is produced from the decoder conditioned on the previous ground truth event $y_{j-1}$, the source context $c_j$, and the target hidden state  $s_{j}$ using Softmax.

The output of a GRU is a $|y|$-dimensional tensor $y$ which represents the probability distribution of each network event element of $x$ over the $|y|$ classes. The GRU is trained over by defining a loss function $L(GRU_\theta(x), y)$ minimized iteratively through backpropagation by varying parameter $\theta$.

\section{Adversary learning- System model} \label{systemmodel}

An IoT device uses protocols such as  UDP, TCP, HTTP, TLS, DNS, DHCP, ARP, and ICMP  while to upload the data on the data server. 
IoT devices can collaborate by using a high range transmission antenna to exchange data. Since collaborating IoT devices are only using  high range channel, therefore, they can  upload the data of distant location as of their location.
The network traffic comprises events of the packets exchanged between the application gateway (AG) and the server on the cloud. This network traffic can be logged onto the application gateway for AI analytics \cite{robyns2017noncooperative}. This traffic includes packets exchanged by collaborating devices also.

AdLIoTLog uses sequences of the packet events of the protocols and subsequences of the packet events logged of each protocol. We model the AdLIoTLog using data with attack and the data without attack for comparison purpose. This model can be trained without actually sharing the data \cite{nguyen2021federated}. Initially, a global model is aggregated thereafter local model updates and provides local model updates to an aggregator. An aggregator combines all local model updates ($AG_1$  with $AG_2$)  and construct a new global  model ($AG_1$  with $AG_2$ and $\ldots$  $AG_n$). Edge devices query the aggregator for any adversary in the network \cite{wang2021milvus}.

\subsection{Model Construction}

The GRU RNN model keeps track of dependencies among elements in the sequences and therefore in the problem of predicting sequences of network events, the GRU is set to learn the set of pairs $(x, y)$ where $x$ is an input sequence of events and $y$ is the expected next sequence of network events.

Let $S$ be a set of  $p$ malicious  nodes represented by $m_1, m_2, .. m_p$. Let $S_1$ is the set of events of  $m_1$ malicious nodes and  $S_p$ is the set of events of $m_p$ malicious nodes. The node $m_1$ perform  $l$ sequence of events\\ ${e_{11} \rightarrow e_{12}, e_{12} \rightarrow e_{13} \ldots  e_{1{l-1}} \rightarrow e_{1l}} $. \\The node $m_p$ perform  $t$ sequence of events \\ ${e_{p1} \rightarrow e_{p2}, e_{p2} \rightarrow e_{p3} \ldots  e_{p{t-1}} \rightarrow e_{pt}} $ 

Therefore it is required to learn a function that can be used for any given source malicious events of ${m_s}$ to predict the targeted coordinated malicious events of ${m_t}$. AdIoTLog collects IoT log over the LAN therefore AdIoTLog comprises of  let $m_1, m_2, .. m_p$  nodes over one application gateway say $AG_1$ while $n_1, n_2, .. n_q$ nodes over another application gateway say $AG_2$.  AdIoTLog computes the probability of possible events in the sequence  P((${m_p: e_{p1} \rightarrow e_{p2}, e_{p2} \rightarrow e_{p3} \ldots  e_{p{l-1}} \rightarrow e_{pl}} $)  $\rightarrow$   (${n_q: e_{q1} \rightarrow e_{q2}, e_{q2} \rightarrow e_{q3} \ldots  e_{q{l-1}} \rightarrow e_{ql}} $)). 

\section{Algorithm} \label{sec:algo}

Our method AdIoTLog aggregator combines all local model updates of $AG_i$  with $AG_j$.  AdIoTLog  input the network events sequences of $AG_i$  and $AG_j$ to the encoder. If direct communication in the IoT network is allowed  then also IoT  nodes across the $AGs$ will not be able to communicate due to low transmission power channel. The malicious IoT devices communicate over the hidden link layer channel. Communications over a "hidden" channel include data packets as well as control packets, such as ARP packets, TCP/UDP packets, and other types of packets. The trace-driven event log comprises the communication across the IoT network.

The sequence to sequence model can map sequences of varying lengths to each other.  Encoder track every output and hidden state such as when a network sequence of size 100 is input with 256 hidden size, it produces encoder output tensor of size (100, 256) and final hidden state tensor of 256 size.

 The decoder uses the last output or the last hidden state of the encoder. The attention helps the decoder network to compute attention weights and then these weights are multiplied by encoder output vectors to create a weighted encoded vector that contains information about the input network event sequence. 
 We used max sentence length of 100 words to train the attention layer. The concept of using target outputs as next input is called teacher forcing ~\cite{lamb2016professor} that helps to converge the training process faster. AdIoTLog  used the teacher forcing algorithm randomly with a probability of 0.5. 

Network loss is computed in AdIoTLog  based on decoder output and target tensor. Network weights for both encoder and decoder are optimized using stochastic gradient descent (SGD) optimizer using a learning rate in range of 0.01-0.0001. We stored loss after every 100 steps to track if the network is learning.

If $A$ is the set of events with the top high probabilities from IoT nodes with same $AG$ then the AdIoTLog  reports nodes to be benign otherwise nodes are reported anomalous.  AdIoTLog takes different IoT traffic patterns depending on the protocols used by the IoT device. When two hosts communicate with one another, it does not always indicate malicious activity; however, if those nodes are not in the range, then it indicates malicious behavior, which is modelled as nodes in two distinct AGs.

The main aim of our method is to feed the higher context, i.e. splitting the input text into contextual content to increase the model output probability distribution so that it matches with the probability distribution of the ground truth values. The TCP packet, UDP packet, and HTTP packet were considered in different contexts. This potentially can reduce the gap between training and inference by training the model to handle the situation, which will appear during test time. 

\begin{figure*}[h]
	\centering
	
	\includegraphics[width=2.5in, height=2in]{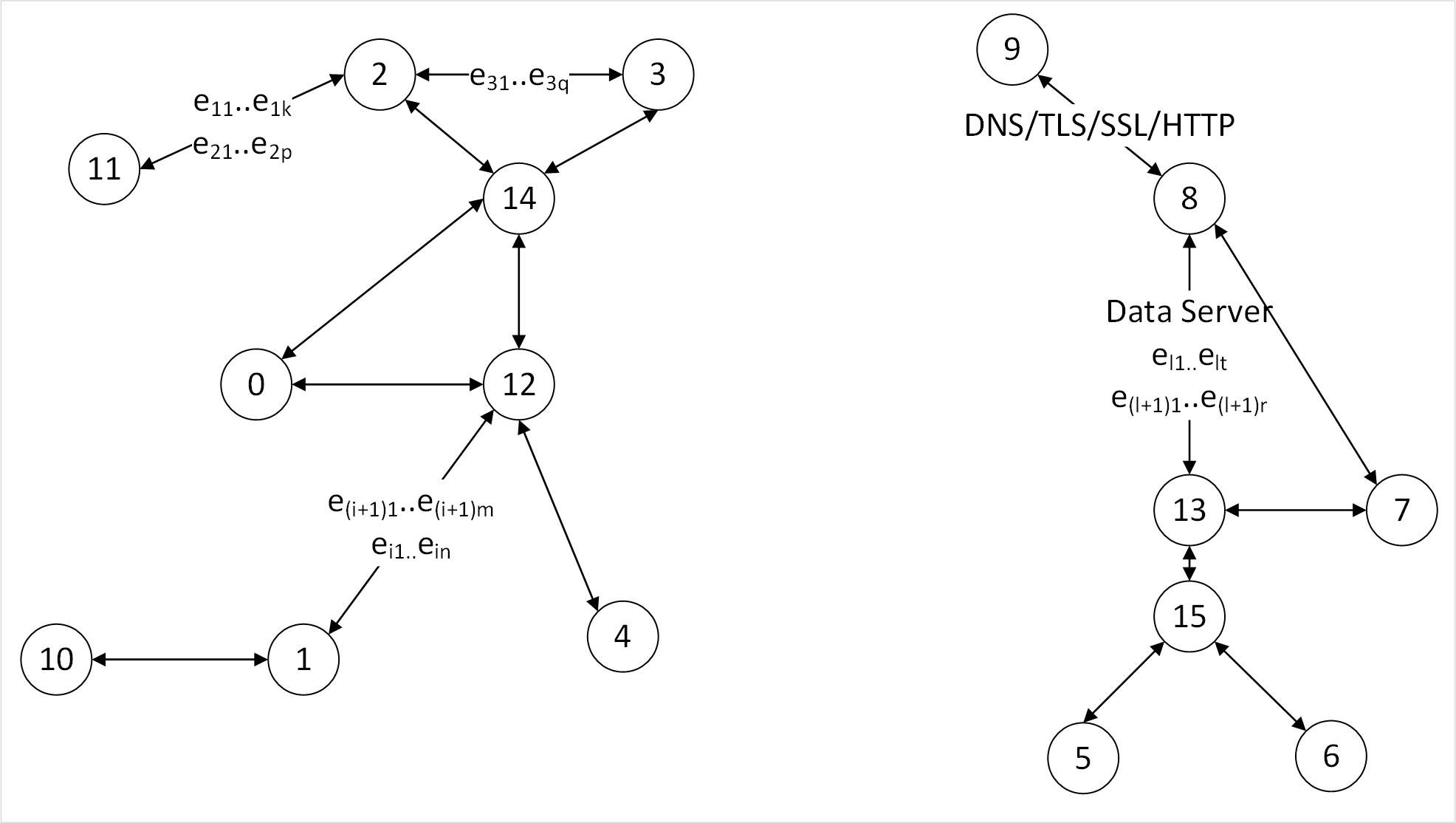}
	\vspace*{0.1cm}
	\includegraphics[width=2.5in, height=2in]{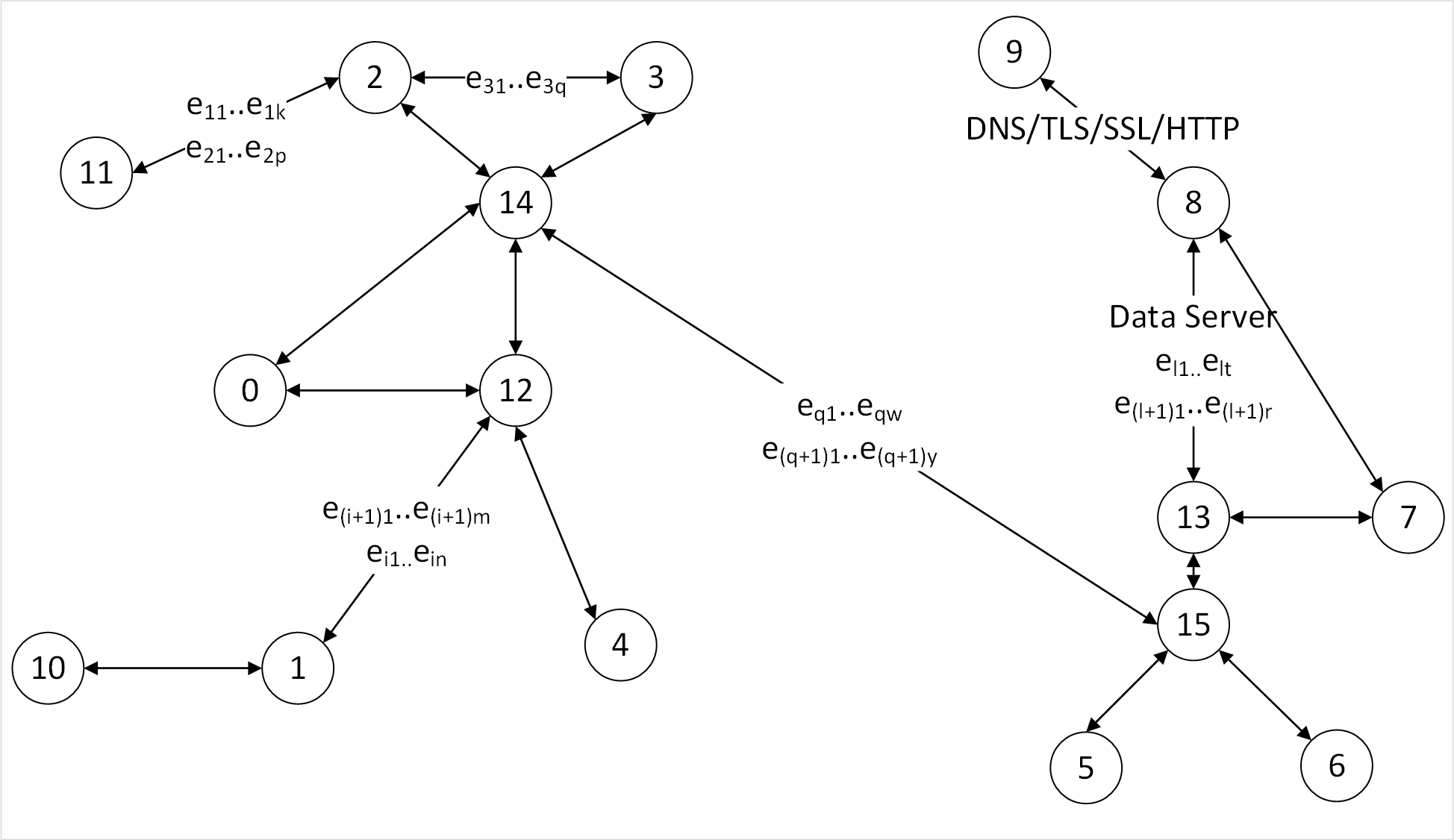}\par\medskip
	
	\caption{(a) IoT network without collaborating nodes in ns-2 (b) IoT network with collaborating nodes in ns-2}
	\label{fig:EventModel}
\end{figure*}

\section{Experiment Setup, Results, and Analysis} 
\label{sec:evaluation}

The GRU RNN model is available in PyTorch. For our experiments, we used Intel 4.7 GHz i7-8700K, 8 GB GTX 1080 with 2560 CUDA cores, and 64 GB Dual Channel DDR4 at 2400 MHz to run the  PyTorch library for GRU  model. The various hyperparameters are explained in Table \ref{table:hyper}.

\subsection{Training Data}
We used ns-2 network simulator dataset to verify the proposed detection of collaborating nodes.  We explain the training process that includes preparing data and training the sequence to sequence network. 

\subsubsection{The dataset prepared using Network Simulator}

The training dataset in ns-2 was created with 16 nodes.  Figure \ref{fig:EventModel} shows two scenarios (a) IoT network without collaborating nodes and (b) IoT network with collaborating nodes. Node pairs were simulated as IoT device and data server. For example, node pair (14, 2) was simulated for node 14 to upload data to node 2. In the first case, when there is no collaboration, node 14 will upload the data to node 2. In the second case, when nodes can collaborate using hidden channel, node 14 will upload the data to node 15.  Eight UDP communication node pairs were used with a 1500 byte packet at the rate of 1 Mbps to generate the data. One pair (14,15) was collaborating adversary nodes which means 12.5\%  of simulated traffic constitutes the attack. AI analytics Sequence-to-Sequence model can remember the good events and collaborative events, therefore, results in detecting the malicious events of the network.

\subsubsection{ Training Sequence-to-Sequence network} 
\begin{figure}
	\centering
	\includegraphics[width=2.5in, height=1.5in]{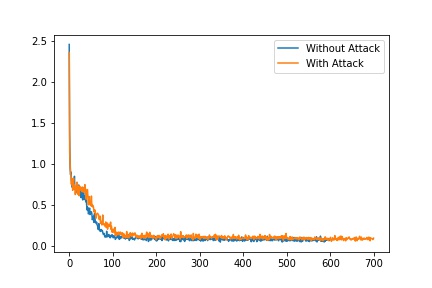}\hfil
	
	\caption{Variation of model training (NLL Loss) x 100 iteration}
	\label{subfig2:shortform}
\end{figure}

To interface ns-2 trace file to RNN model, we need a tensor pair.  A tensor indexes IoT network log vocabulary for the input of the model.  A tensor pair was prepared by including network events input tensor and network events target tensor. The logged network events were paired following the order of timestamps of network events one after the other. The input file included 12,236 network sequence pairs with 4170 unique elements that comprise different types of packets, protocols, sequence numbers, and flags.  The combined ns-2 trace files of both setups - networks with hidden channel and network without hidden channel were input to the model. To compare the two scenarios, GRU RNN was also trained without adversary node. Figure~\ref{subfig2:shortform} shows variation of model training (Negative Log Likelihood Loss) with moving average of over 100 iterations. Table \ref{table:hyper} shows hyperparameters used in the training process.

\subsection{Results and Analysis}
\label{sec:results}

We define the following performance metric based on BLEU score \cite{papineni2002bleu}.  The BLEU score was computed by comparing the predicted network event sequence with the ground truth network sequence using 1-gram (single words). It is 100 if predicted sequence is exactly similar to ground truth sequence. We define accuracy of the model that is based on the number of testing pairs as follows:\\ 

${\bf Accuracy }  = \frac{sum~ of ~BLEUscore(TestingPairs) }{len(TestingPairs) }$ \\

Let ${tp_1, tp_2, ..., tp_n}$ are testing pairs and their respective bleu scores are ${b_1, b_2, ..., b_n}$ then accuracy of model output will be  $ \frac{\sum_{i=1}^{n} b_i}{n} $.

\begin{table}[htp]
	\caption{Hyper parameters of model}
	\label{table:hyper}
	\begin{center}
		\begin{tabular}{|c|p{1cm}|p{.9cm}|p{1cm}|p{1.2cm}|p{1cm}|}
			\hline
			Network & No of hidden layer & No of\linebreak iterations & Learning rate & Hidden Layer size & optimizer  \\ [0.1ex] 
			
			\hline
			IoT ns-2 & 1   &   70,000 & 0.01-0.0001 & 256 &  SGD\\
			\hline
			
		\end{tabular}
	\end{center}
	\vspace{-4mm}
\end{table}

\begin{table*}[htp]
	\caption{Comparison of Model Output under collaborative-attack and non-collaborative attack scenarios}
	\label{table:modelp}
	\begin{center}
		\begin{tabular}{ | p{1.3cm} | p{1.2cm}|p{3.5cm}|p{3.5cm} || p{1.5cm} | p{2cm} | } 
			\hline
			Network & Size of set $A$ & Node tuple in $A$ in data without attack  (node, actual data server, pred data server) & Node tuple in data with attack  (node, actual data server, pred data server) &Accuracy with collaborative attack & Accuracy without collaborative attack\\ [0.1ex] 
			
			\hline
			IoT ns-2 & 5   &  (6,15,15), (3,2,14), (6,15,15), (2,3,3), (8,9,9)    & (14,2,15), (0,12,12), (2,11,14), (0,12,12), (15,5,6) & 89-95\% & 91-98\%\\
			\hline
		\end{tabular}
	\end{center}
	\vspace{-4mm}
\end{table*}

Table~\ref{table:modelp} shows model performance comparison of model output under collaborative-attack and non-collaborative attack scenarios. The experimental results show accuracy of 89-95\% in case of collaborative attack and 91-98\% in case of non-collaborative attack.

Next, we explain the findings of the experiments based on model performance. The experiments were designed to answer the following questions:

\paragraph{What is the performance of the GRU RNN-based models when the input network events use sequences of the packet events of the protocols  such as TCP, UDP, and HTTP and subsequences of the packet events like sequence number,  IP addresses, and window scale option logged for each protocol}

In simulator trace files,  the values of features like sequence numbers, flags values, and IP addresses were simulated values different from the format used by TCP/IP model on a network. For example, IP address in dotted format (e.g.  192.168.1.1) was replaced by a node number (e.g. 14).  The generated sequence numbers were much easier to keep track of relatively small, predictable numbers rather than the actual numbers. Acknowledgement numbers were also not very random. We observed the high accuracy of GRU RNN in predicting the network events for simulator data in comparison to dumped TCP/IP model output on a IoT network.

\paragraph{How is the subset $A$ of  $x$ with top probabilities over $|y|$-dimensional tensor $y$ in $|y|$ classes changed using the data under collaborative attack (shown in Figure \ref{fig:EventModel}a) and non attack scenario (shown in Figure \ref{fig:EventModel}b)}

When the results of both models (a) with attack data and (b) without attack data were compared with each other on the set $A$, we observed that the accuracy of predicted network subsequences was less in the presence of attack. The collaborating nodes were connected through hidden channel, therefore the communication of adversarial node in $N_1$ with collaborating adversarial node in $N_2$ superseded in comparison to other nodes in their respective networks $N_1$ and $N_2$.  The model was biased on predicting the collaborating class in the $|y|$ classes.
In other words, in AdIoTLog under attack scenario,
P((${m_p: e_{p1} \rightarrow e_{p2}, e_{p2} \rightarrow e_{p3} \ldots  e_{p{l-1}} \rightarrow e_{pl}}$) $\rightarrow$  \\
(${n_q: e_{q1} \rightarrow e_{q2}, e_{q2} \rightarrow e_{q3} \ldots  e_{q{l-1}} \rightarrow e_{ql}} $)) 
was high most of the times for collaborating class.   If we present Node tuple in $A$ as (node, actual data server, pred data server) shown in Table \ref{table:modelp}, for tuple (14, 2, 15), the actual node event was by 2 while model predicted node 15 rather than node 2 for source 14.

\paragraph{How the performance of model varied when used for scenario with collaborating malicious nodes}
In simulator generated log, there were sufficient instances needed for model learning.  The model performance reduced in the presence of the attack. Features were much easier to keep track because of relatively small, predictable numbers rather than the actual numbers. We observed that when there is less data to train we get more iteration however when there is more data, number of iteration are less. GRU RNN can predict the event when trained on even small data with context. 

Figure~\ref{subfig2:shortform} shows variation of model training (Negative Log Likelihood Loss) with moving average over 100 iterations.

\section{Background}


Wu et al. \cite{wu2019bigdata}   presented a method for analyzing the bigdata collected through tools e.g. Hadoop, Mapreduce, Hive and others based on seq2seq- predictive models. Rather than input the seq vector from the bigdata components, the authors labeled the data of each component, to get the embedding vector, and subsequently, input labeled vector to attention matrix. Finally, the predicted vector is obtained using target information. In contrast, in our proposed method we split input vector on the basis of context of protocol which is processed by encoder and subsequently processed by decoder incorporating attention using target vector. 

The model developed by  Shen et al. \cite{shen2018tiresias} for collaborative nodes trying to use vulnerabilities of intrusion protection system also used RNN. The negative shift in model prediction was used to detect the attack, however, the attack considered was not over distributed machines, also they did not compare output of system under attack and without attack scenarios. They varied the length of input sequence to study the performance of RNN with increasing length of input sequence. Although, we observed that RNN shows better results if delimited using context rather than by arbitrarily varying the length of sequences input to the model. 

The model developed by Almiani et al. \cite{almiani2019deep} for intrusion detection system in an IoT network was trained on the NSL-KDD dataset for different types of attacks.  
Amanullah et al. \cite{amanullah2020deep} studied and presented the deep learning technologies for IoT security.
The growth in use of deep learning models for security shows its promising applicability on IoT logs. 

\section{Conclusion}
\label{sec:con}
In this paper, we studied performance of GRU RNN model on the traffic log of an IoT network with and without adversary nodes. The adversary nodes are assumed to collaborate to cripple the data to be uploaded.  We found that adversary nodes can be detected without considering any additional events rather it is required to log the network traffic.  The log can be analyzed by AI algorithms to detect the adversary nodes in the network. 

\small

\bibliographystyle{unsrt}
\bibliography{reference}

\end{document}